\documentclass[letterpaper, 10 pt, conference]{ieeeconf}
\usepackage{etex}

\IEEEoverridecommandlockouts                              

\usepackage{times}
\usepackage{latexsym,amsfonts,amsbsy,amssymb}
\usepackage{amsmath}
\usepackage{graphicx}
\usepackage{cite}
\usepackage{algorithm}
\usepackage{algorithmic}
\usepackage{multirow}
\usepackage{gensymb}
\usepackage{mathrsfs}
\usepackage{stmaryrd}
\usepackage{tabularx,booktabs}
\usepackage[table]{xcolor}
\usepackage[curve]{xypic}
\usepackage{subfig}
\usepackage{caption}
\usepackage{color}
\usepackage{tikz}
\usepackage{flushend}
\usetikzlibrary{automata,positioning,shapes,arrows}

\newtheorem{definition}{Definition}

\newtheorem{problem}{Problem}

\usepackage{float}

\usepackage[algo2e]{algorithm2e}

\begin{document}
\title{\LARGE \bf Distributed Communication-aware Motion Planning for Multi-agent Systems from STL and SpaTeL Specifications  \thanks{This work was supported by the National Science Foundation (NSF-CNS-1239222 and NSF- EECS-1253488)}}
\author{Zhiyu~Liu,
	    Bo~Wu,
        Jin~Dai, 
        and Hai~Lin 
\thanks{Z. Liu, B. Wu, J. Dai,  and H. Lin are with the Department of Electrical Engineering, University of Notre Dame, Notre Dame, IN 46556, USA (e-mail: zliu9@nd.edu; bwu3@nd.edu; jdai1@nd.edu; hlin1@nd.edu).}}
\date{}

\maketitle
\thispagestyle{empty}
\pagestyle{empty}

\begin{abstract}
In future intelligent transportation systems, networked vehicles coordinate with each other to achieve safe operations based on an assumption that communications among vehicles and infrastructure are reliable. Traditional methods usually deal with the design of control systems and communication networks in a separated manner. However, control and communication systems are tightly coupled as the motions of vehicles will affect the overall communication quality. Hence, we are motivated to study the co-design of both control and communication systems. In particular, we propose a control theoretical framework for distributed motion planning for multi-agent systems which satisfies complex and high-level spatial and temporal specifications while accounting for communication quality at the same time. Towards this end, desired motion specifications and communication performances are formulated as signal temporal logic (STL) and spatial-temporal logic (SpaTeL) formulas, respectively. The specifications are encoded as constraints on system and environment state variables of mixed integer linear programs (MILP), and upon which control strategies satisfying both STL and SpaTeL specifications are generated for each agent by employing a distributed model predictive control (MPC) framework. Effectiveness of the proposed framework is validated by a simulation of distributed communication-aware motion planning for multi-agent systems.
\end{abstract}

\section{Introduction}
Multi-agent systems like intelligent transportation systems and smart cities have emerged as a hot research topic in the interdisciplinary study of control theory, robotics and computer science due to their wide applications in both academic and industrial domains. Being one of the fundamental research problems in this context, motion planning and control of multi-agent systems has drawn a considerable amount of research interest in recent years. Motion and control strategies are developed for various global/local coordination purposes, see e.g. \cite{cao2013overview} and the references therein.

Recent theoretical and technological developments have enhanced the application of {\it formal methods} to address complex and high-level motion planning objectives. Formal languages, such as linear temporal logic (LTL) and computation tree logic (CTL) \cite{baier2008principles}, provide a concise formalism for specifying and verifying desired logic behavior of dynamical systems \cite{kloetzer2008fully}. To pursue satisfaction of temporal specifications, a vast majority of research efforts has been devoted to {\it abstraction-based} approaches (see e.g. \cite{belta2007symbolic,kloetzer2010automatic,fainekos2009temporal} and the references therein), where the temporal logic formula specification is translated into an automata representation whose accepting runs correspond to satisfaction of the formula, while the environment of the system is also abstracted into a finite transition diagram. Based on these abstractions, algorithms are derived for verification and synthesis of discrete controllers that drive the system to satisfy the specification. Despite their success in the correct-by-construction design of controllers, abstraction-based approaches suffer from the ``state explosion" issues as both the synthesis and abstraction algorithms scale at least exponentially with the dimension of the discretized configuration space \cite{kloetzer2008fully}, rendering such approaches impractical for large-scale multi-agent systems. Furthermore, LTL and/or CTL specifications require only the order of events that should be executed by the system, whereas temporal distance between them is often neglected. Many attempts have been made to multi-agent systems to ease the computational burden while obeying temporal logic specifications. On the one hand, approaches that do not require abstractions are proposed, such as sampling-based approach \cite{karaman2009sampling} and mixed integer linear programming techniques \cite{richards2002aircraft,karaman2011linear,wolff2014optimization}. On the other hand, model predictive control (MPC) schemes are introduced to reduce the formal synthesis problem into a series of smaller optimization problems of a shorter horizon. Wongpiromsarn et al. \cite{wongpiromsarn2012receding} utilized MPC to accomplish motion planning for LTL specifications, whereas Tumova and Dimarogonas \cite{tumova2016multi} applied MPC for coordination of multi-agent systems. Kuwata and How \cite{kuwata2011cooperative} integrated MILP formalism with MPC techniques to pursue distributed trajectory optimization; nevertheless, satisfaction of more complex specifications were not studied.

It is worth pointing out that aforementioned results assumed perfect inter-agent communication, which turned out to be oversimplified in practice, since communication quality of service (QoS) is crucial in maintaining multi-agent coordination. In intelligent transportation systems, communications among vehicles and infrastructure are not always reliable. To pursue the co-optimization of motion and communication, Grancharova et al. \cite{grancharova2015uavs} proposed a trajectory planning scheme for agents subject to communication capacity constraints by applying distributed MPC. The same approach is utilized in \cite{grotli2012path} for motion planning of multi-agent systems under radio communication constraints while agents are treated as relay nodes where communication channels are modeled as disk model which assumes communication is off beyond a certain threshold. Yan and Mostofi \cite{yan2013co} modeled the communication channel between a robot and a base station as a Gaussian process with fading and shadowing effects; however, the optimization was performed only with respect to the robot's motion velocity, transmission rate and stop time, while the robot was assumed to travel along a pre-defined trajectory.  Nevertheless, communication models presented in the previous work are either too complicated for multi-agent systems or oversimplified in practice.

Motivated by the aforementioned concerns, we focus on construction of local controller for multi-agent systems such that certain motion requirements are fulfilled in the presence of communication capacity constraints in this paper. Given multiple agents moving in a shared environment with communication base stations, our design objective is to drive each agent to satisfy its own motion specifications, while unsafe zone, collision and poor communication QoS are avoided. Towards this end, we use signal temporal logic (STL) \cite{maler2004monitoring,raman2014model} formulas to describe local motion and safety requirements for each agent, and spatial-temporal logic (SpaTel) \cite{haghighi2016robotic} formulas to represent global safety and  communication QoS requirements, based on which an MILP formalism is established to solve not only the joint motion-communication co-optimization problem, but synthesizes collision-avoiding motion controllers as well. Note that different from our previous work \cite{liuacc}, we present a distributed synthesis framework by employing MPC such that appropriate controllers can be synthesized locally.



%
%

The rest of this paper is organized as follows. In Section II, we briefly introduce necessary preliminaries of STL and SpaTel. We then formally present the communication-aware motion planning problem from STL-SpaTel specifications in Section III. In Section IV, we propose the MILP encoding schemes for STL and SpaTel specifications. Based on these MILP constraints, we exploit a distributed MPC strategy in Section V to synthesize local controllers for each agent. Simulation examples are presented in Section VI for validating our proposed framework. Section VII concludes the paper.

\section{Preliminaries}

\subsection{Agent Models}
The multi-agent system under consideration in this paper consists of  $P$ agents with unique identities $\mathcal{P}=\{1,2, \ldots, P\}$ which perform their missions in a shared 2-D environment. For each $i\in \mathcal{P}$, the agent dynamics is dominated by the linear dynamics of the following form
\begin{equation}
\dot x_i(t)=Ax_i(t)+Bu_i(t),
\label{dynamic}
\end{equation}
where $x_i\in\mathbb{R}^4$ is the state of agent $i$ with $x_i=[p_i^T\quad v_i^T]^T$, where $p_i, v_i\in\mathbb{R}^2$ are the position and velocity of the agent, respectively; $u_i=[u_{i,1}\quad u_{i,2}]^T\in \mathcal{U}\subseteq \mathbb{R}^2$ is the local admissible control inputs, and $x_i(0)=x_{i,0}\in \mathbb{R}^4$ is the initial state. $(A,B)$ is a controllable pair with proper dimensions. The environment $\mathcal{X}$ is given by a large convex polygonal subset of the $2$-D Euclidean space $\mathbb{R}^2$. Let $\mathcal{X}_{obs}\subseteq \mathcal{X}$ be the regions in the environment occupied by polygon obstacles. $\mathcal{X}_{free}=\mathcal{X}\setminus\mathcal{X}_{obs}$ denotes the obstacle-free working space for the multi-agent system.

To run the distributed communication-aware motion planning in an online manner, we borrow the idea from \cite{raman2015reactive} and assume that the agent dynamics (\ref{dynamic}) admits a discrete-time approximation of the following form, given an appropriate sampling time $\Delta t>0$:

\begin{equation}
x_i(t_{k+1})=A_dx_i(t_{k})+B_du_i(t_{k}),
\label{agentdynamic}
\end{equation}
where $k\in\mathbb{N}$ is the sampling index and $\Delta t$ is selected such that $(A_d, B_d)$ is controllable. The sampling is uniformly performed, i.e., for each $k>0$, $t_{k+1}-t_k=\Delta t$ and we use $[a,b]$ as an abbreviation for the set $\{a,a+1,\ldots,b\}$.

Given $x_{i,k}\in\mathbb{R}^2$ and $\bf{u}_i^H\in \mathcal{U}^\omega$, $i\in\mathcal{P}$, a (state) {\it run} ${\bf x}_i^H= x_{i,k}x_{i,k+1}x_{i,k+2}\ldots x_{i,k+H-1}$ generated by agent $i$'s dynamics (2) with control input $\bf {u_i^H}$ is a finite sequence obtained from agent $i$'s state trajectory, where $x_{i,k}=x_i(t_k)\in\mathbb{R}^4$ is the state of the system at time index $t$, and for each $k\in\mathbb{N}$, there exists a control input $u_{i,k}=u_i(t_k)\in \mathcal{U}$ such that $x_i(t_{k+1})=A_dx_i(t_{k})+B_du_i(t_{k})$. Under MPC framework with planning horizon $H$ (cf. Section III), given a local state $x_{i,k}$ and a sequence of local control inputs ${\bf u}^H_i = u_{i,k}u_{i,k+1}u_{i,k+2}\ldots u_{i,k+H-1}$, the resulting horizon-$H$ run of agent $i$, ${\bf x}_i(x_{i,k},{\bf u}^H_i)=x_{i,k}x_{i,k+1}x_{i,k+2}\ldots x_{i,k+H-1}$ is unique.

\subsection{Communication}
Typically, agents within the team may be assigned with different roles and responsibilities, and inter-agent communication is required to ensure proper coordination between them for safety and efficient mission execution. Furthermore, reliable communication is also needed between the agents and base stations to collect the sensed environment data and let base stations provide global information for agents, which is essential for distributed algorithms. Therefore in this paper we explicitly consider communication as an optimization objective.

\begin{figure}[h]
	\centering
	\includegraphics[width=0.6\linewidth]{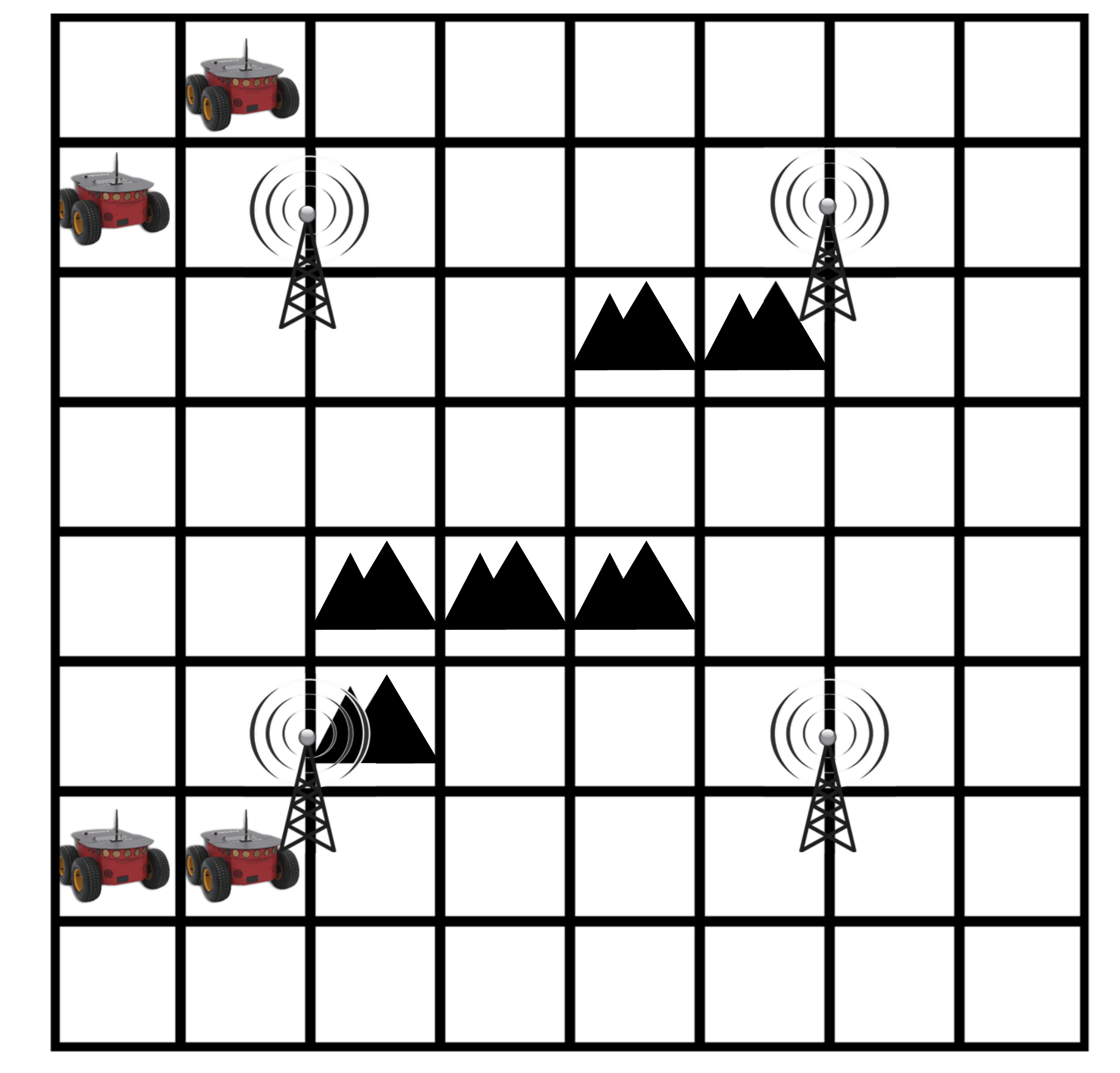}
	\caption{The gridded environment}
	\label{comm}
\end{figure}

As shown in Fig. \ref{comm}, we assume the environment is a gridded square world. There are four base stations, one at the center of each quadrant. To reduce the interference and increase the number of agents that can be served, similar to cellular communication, each base station is equipped with four $90\degree$ sector antenna systems to fully cover each quadrant. The quality of service (QoS) of inter-agent communication and between the individual agent and the base station is assumed to be subject to the path loss and the shadowing effect \cite{molisch2012wireless} due to the mountains.

\subsection{Signal Temporal Logic}
We consider STL formulas which are defined recursively as follows.

\begin{definition}[STL Syntax]\rm
STL formulas are defined recursively as:
$$
\varphi::={\rm True}|\pi^\mu|\neg\pi^{\mu}|\varphi\land\psi|\varphi\lor\psi|\Box_{[a,b]} \psi | \varphi\sqcup_{[a,b]} \psi
$$
\end{definition}
where $\pi^\mu$ is an atomic predicate $\mathbb{R}^n\to\{0,1\}$ whose truth value is determined by the sign of a function $\mu:\mathbb{R}^n\to\mathbb{R}$, i.e., $\pi^\mu$ is true if and only if $\mu({\bf x})>0$; and $\psi$ is an STL formula. The ``eventually" operator $\Diamond$ can also be defined here by setting $\Diamond_{[a,b]} \varphi={\rm True}\sqcup_{[a,b]} \varphi$.

The semantics of STL with respect to a discrete-time signal $\bf x$ are introduced as follows, where $({\bf x},t_k)\models \varphi$ denotes for which signal values and at what time index the formula $\varphi$ holds true.
\begin{definition}[STL Semantics]\rm
The validity of an STL formula $\varphi$ with respect to signal $\bf x$ at time $t_k$ is defined inductively as follows:
\begin{enumerate}
\item $({\bf x},t_k)\models \mu$, if and only if $\mu(x_k)>0$;
\item $({\bf x},t_k)\models \neg\mu$, if and only if $\neg(({\bf x},t_k)\models \mu)$;
\item $({\bf x},t_k)\models \varphi\land\psi$, if and only if $({\bf x},t_k)\models \varphi$ and $({\bf x},t_k)\models \psi$;
\item $({\bf x},t_k)\models \varphi\lor\psi$, if and only if $({\bf x},t_k)\models \varphi$ or $({\bf x},t_k)\models \psi$;
\item $({\bf x},t_k)\models \Box_{[a,b]}\varphi$, if and only if $\forall t_{k'}\in[t_k+a,t_k+b]$, $({\bf x},t_{k'})\models \varphi$;
\item $({\bf x},t_k)\models \varphi\sqcup_{[a,b]}\psi$, if and only if $\exists t_{k'}\in[t_k+a,t_k+b]$ such that $({\bf x},t_{k'})\models \psi$ and $\forall t_{k''}\in[t_k,t_{k'}]$, $({\bf x},t_{k''})\models \varphi$.
\end{enumerate}
\end{definition}

A signal ${\bf x}= x_0x_1x_2\ldots$ satisfies $\varphi$, denoted by $\bf x\models\varphi$, if $({\bf x},t_0)\models\varphi$.

Intuitively, $\bf x\models \Box_{[a,b]}\varphi$ if $\varphi$ holds at every time step between $a$ and $b$, ${\bf x}\models \varphi\sqcup_{[a,b]}\psi$ if $\varphi$ holds at every time step before $\psi$ holds, and $\psi$ holds at some time step between $a$ and $b$, and ${\bf x}\models \Diamond_{[a,b]}\varphi$ if $\varphi$ holds at some time step between $a$ and $b$.

An STL formula $\varphi$ is {\it bounded-time} if it contains no unbounded operators. The bound of $\varphi$ can be interpreted as the horizon of future predicted signals $\bf x$ that is needed to calculate the satisfaction of $\varphi$.

\subsection{Spatial Temporal Logic}
SpaTeL is defined by the combination of STL and Tree Spatial Superposition Logic (TSSL) where STL is responsible for describing the temporal properties and TSSL is related to spatial properties \cite{haghighi2015spatel}. Prior to define the syntax and semantics of SpaTeL, we first introduce quad transition systems (QTS) with a quad tree data structure to model spatial characteristics.

A QTS \cite{haghighi2015spatel} is defined through a quad tree structure by dividing environment into four quadrants recursively. It is defined as a tuple $Q(t)=(\mathcal{V}, \mathcal{E}, v_0, V_f, \mu, \mathcal{L}, l)$, where $\mathcal{V}$ is the set of nodes $v_i$. $\mathcal{E} \subset \mathcal{V} \times \mathcal{V}$ is the set of directed transitions. $v_2$ is a child of $v_1$ if $(v_1,v_2) \in \mathcal{E}$. $v_0$ is the root of the tree (the only node which is not a child of another node). $V_f$ is the set of leaves (nodes without children). $\mu : \mathcal{V} \times \mathbb{R}_{\geq 0} \rightarrow \mathbb{R}^{+}$ is the valuation function assigning each node a real positive number. $\mathcal{L}$ is a finite set of labels. $l : \mathcal{E} \rightarrow \mathcal{L}$ is the labeling function which maps each edge to a label. A labeled path of a QTS is defined as a function which maps a node to a set of infinite sequences of nodes:
\begin{equation}
\lambda^\mathcal{B} := \begin{Bmatrix}
(v_0,v_1,v_2,\dots,\bar{v}_f) | (v_i,v_{i+1})\in \mathcal{E}, l(v_i,v_{i+1}) \in \mathcal{B}
\end{Bmatrix},
\end{equation}
where $\mathcal{B}\subseteq \mathcal{L}$; $\bar{v}_f$ denotes infinite repetitions of leaf node $v_f$; $i\in \mathbb{N}_{\geq 0}$.
A trace corresponding to a trajectory traveling in the divided environment is defined as a function ${\bf q} :\begin{Bmatrix}
0,\dots,T
\end{Bmatrix} \rightarrow Q$.

\begin{figure}
	\centering
	\begin{tabular}{ | l | l | l | l | l | l | l | l| }
		\hline
		6\cellcolor[gray]{0.9} & 4\cellcolor[gray]{0.7} & 3\cellcolor[gray]{0.6} & 2\cellcolor[gray]{0.5} & 2\cellcolor[gray]{0.5} & 3\cellcolor[gray]{0.6} & 3\cellcolor[gray]{0.6} & 2\cellcolor[gray]{0.5}\\ \hline
		4\cellcolor[gray]{0.7} & 4\cellcolor[gray]{0.7} & 4\cellcolor[gray]{0.7} & 3\cellcolor[gray]{0.6} & 3\cellcolor[gray]{0.6} & 4\cellcolor[gray]{0.7} & 6\cellcolor[gray]{0.9} & 3\cellcolor[gray]{0.6}\\ \hline
		3\cellcolor[gray]{0.6} & 4\cellcolor[gray]{0.7} & 4\cellcolor[gray]{0.7} & 2\cellcolor[gray]{0.5} & 0\cellcolor{red} & 0\cellcolor{red} & 4\cellcolor[gray]{0.7} & 3\cellcolor[gray]{0.6}\\ \hline
		2\cellcolor[gray]{0.5} & 3\cellcolor[gray]{0.6} & 3\cellcolor[gray]{0.6} & 2\cellcolor[gray]{0.5} & 1\cellcolor[gray]{0.3} & 2\cellcolor[gray]{0.5} & 3\cellcolor[gray]{0.6} & 2\cellcolor[gray]{0.5}\\ \hline
		2\cellcolor[gray]{0.5} & 3\cellcolor[gray]{0.6} & 0\cellcolor{red} & 0\cellcolor{red} & 0\cellcolor{red} & 3\cellcolor[gray]{0.6} & 3\cellcolor[gray]{0.6} & 2\cellcolor[gray]{0.5}\\ \hline
	    3\cellcolor[gray]{0.6} & 4\cellcolor[gray]{0.7} & 0\cellcolor{red} & 2\cellcolor[gray]{0.5} & 3\cellcolor[gray]{0.6} & 4\cellcolor[gray]{0.7} & 4\cellcolor[gray]{0.7} & 3\cellcolor[gray]{0.6}\\ \hline
		4\cellcolor[gray]{0.7} & 4\cellcolor[gray]{0.7} & 4\cellcolor[gray]{0.7} & 3\cellcolor[gray]{0.6} & 3\cellcolor[gray]{0.6} & 4\cellcolor[gray]{0.7} & 6\cellcolor[gray]{0.9} & 3\cellcolor[gray]{0.6}\\ \hline
		6\cellcolor[gray]{0.9} & 4\cellcolor[gray]{0.7} & 3\cellcolor[gray]{0.6} & 2\cellcolor[gray]{0.5} & 2\cellcolor[gray]{0.5} & 3\cellcolor[gray]{0.6} & 3\cellcolor[gray]{0.6} & 2\cellcolor[gray]{0.5}\\
		\hline
	\end{tabular}
	\caption{Matrix C for the environment}
	\label{matrix}
\end{figure}

\begin{figure}
	\centering
	\scalebox{0.7}{
	\begin{tikzpicture}[nodes={circle}, ->]
	\node [circle,draw] {$v_0$}
	child { node  [circle,draw] {$v_1$}
		child { node [circle,draw] {$v_5$} edge from parent node [left,font=\small] {NW} }
		child { node [circle,draw] {$v_6$} edge from parent node [left,font=\small] {NE}}
		child { node [circle,draw] {$v_7$} edge from parent node [right,font=\small] {SW}}
		child { node [circle,draw] {$v_8$}
			child { node [circle,draw] {$v_{f}$} edge from parent node [left,font=\small] {NW}}
			child { node [circle,draw] {$v_{f}$} edge from parent node [left,font=\small] {NE}}
			child { node [circle,draw] {$v_{f}$} edge from parent node [right,font=\small] {SW}}
			child { node [circle,draw] {$v_{f}$} edge from parent node [right,font=\small] {SE}} edge from parent node [right,font=\small] {SE}} edge from parent node [left,font=\small] {NW}}
	child { node [circle,draw] {$v_2$} edge from parent node [left,font=\small] {NE} }
	child { node [circle,draw] {$v_3$} edge from parent node [right,font=\small] {SW}}
	child { node [circle,draw] {$v_4$} edge from parent node [right,font=\small] {SE}};
	\end{tikzpicture}
	}	
	\caption{Partial QTS for Matrix C}
	\label{tree}
\end{figure}
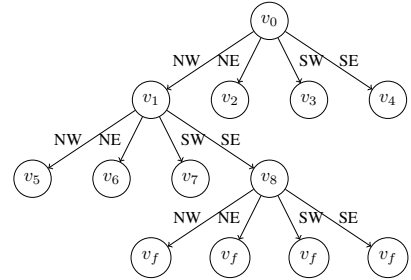
Fig. \ref{matrix} and Fig. \ref{tree} show an example of formulating QTS. Let matrix $C \in\mathbb{R}^{2^{D} \times 2^{D}}$ be the representation of the environment in Fig. \ref{comm} divided by $2^D \times 2^D$ grids and elements in it represent a valuation function $\mu$ for corresponding area, where $D \in\mathbb{N}^+$ is the resolution of the matrix (or the depth of the tree), which is 3 in this case. A QTS can be constructed from $C$ by choosing it as the root node $v_0$ then dividing $C$ into four $2^{D-1} \times 2^{D-1}$ sub-matrices. Each of them is a child of $v_0$ and the directed transitions are stored in $\mathcal{E}$ with  directional labels from the set $\mathcal{L} = \begin{Bmatrix}
NW,NE,SE,SW
\end{Bmatrix}$. One can expand each child with the same method applied on the root node until we have $2^D \times 2^D$ leaf nodes. Valuation function $\mu$ can be defined for each leaf nodes by using the elements in $C$ and for other nodes by the mean of the valuations of its children recursively:

\begin{equation}
\mu(v)=\frac{1}{4}\sum_{(v,v_c)\in\mathcal{E}}\mu(v_c),  \forall v\in \mathcal{V} \setminus V_f,
\label{valfun}
\end{equation}

Based on the definition of QTS, the syntax and semantics of TSSL are well defined in \cite{bartocci2016formal}. With STL and TSSL, one can define SpaTeL as follow \cite{haghighi2015spatel}.

\begin{definition}[SpaTeL Syntax]\rm
	The spatial formulas are defined recursively as:
	$$
	\begin{aligned}
		\psi::=&{\rm True}|m \sim d|\neg\psi|\psi_1\land\psi_2|\exists_B \bigcirc \psi|\forall_B \bigcirc \psi |\\
		 &\exists_B \psi_1U_k\psi_2 | \forall_B \psi_1U_k\psi_2
	\end{aligned}
	$$

where $\sim \in \begin{Bmatrix}
\geq, \leq
\end{Bmatrix}$, $d \in [0,b]$, $b \in \mathbb{R}^+$, $k \in \mathbb{N}^{+}$, $B \subseteq \mathcal{L} $ with $B\neq \emptyset $, and $m \in \mu$. $U_k$ and $\bigcirc$ are until and next operators respectively. The syntax of SpaTeL formulas are defined as follow:
$$
\phi ::= \psi|\neg\phi|\phi_1\land\phi_2|\Box_{[a,b]} \phi|\phi_1\sqcup_{[a,b]} \phi_2.
$$
\end{definition}
As we noticed, SpaTeL is a combination of STL and TSSL by replacing the STL predicates with TSSL formulas.

\begin{definition}[SpaTeL Semantics]\rm
	The validity of an SpaTeL formula $\phi$ with a trace $\bf q$ at time $t_k$ is defined inductively as follows:
\begin{enumerate}
	\item $({\bf q},t_k)\models \psi $, if and only if $({\bf q},v_0,t_k)\models \psi $;
	\item $({\bf q},t_k)\models \neg \phi $, if and only if $\neg(({\bf q},t_k)\models \phi) $;
	\item $({\bf q},t_k)\models \phi_1\land\phi_2$, if and only if $({\bf q},t_k)\models \phi_1$ and $({\bf q},t_k)\models \phi_2$;
	\item $({\bf q},t_k)\models \Box_{[a,b]}\phi$, if and only if $\forall t_{k'}\in[t_k+a,t_k+b]$, $({\bf q},t_{k'})\models \phi$;
	\item $({\bf q},t_k)\models \phi_1\sqcup_{[a,b]} \phi_2$, if and only if $\exists t_{k'}\in[t_k+a,t_k+b]$ such that $({\bf q},t_{k'})\models \phi_2$ and $\forall t_{k''}\in[t_k,t_{k'}]$, $({\bf q},t_{k''})\models \phi_1$;
\end{enumerate}
where $\psi$ is a TSSL formula.
\end{definition}

\subsection{Distributed MPC}
Instead of solving optimization problems for the whole time horizon $T_f$, MPC solves problems within a finite horizon $H<T_f$ starting from current states and provides a finite control inputs $u^H_k$. Only the first control input will be implemented and agents' states will be sampled again. Then the new optimization problem will be computed based on the new states within the finite horizon $H$. MPC has a better performance comparing to other methods in terms of handling model uncertainty and external disturbance \cite{mayne2000constrained}. It also reduces the size of problem by only considering $H$ steps ahead. For distributed MPC, each agent has its own MPC based optimization problem which only considers its own neighbor \cite{kuwata2011cooperative} and therefore makes it much smaller than the centralized problem. By applying distributed MPC, one can deal with larger scale cases\cite{kuwata2007distributed}.

\section{Problem Statement}

\subsection{STL Motion Planning Specifications}
We now proceed to the communication-aware motion planning problems for multi-agent systems. Let us consider a team of $P$ agents conducting motion behavior in the shared environment $\mathcal{X}$, each of which is governed by the discretized dynamics (2). We assign a goal region $\mathcal{X}_{i,goal}$ for agent $i$, $i\in\mathcal{P}$ that is characterized by a {\it polytope} \cite{kloetzer2008fully} in $\mathcal{X}_{free}$, i.e., there exist $M\ge3$ and $a_{i,j}\in \mathbb{R}^2$, $b_{i,j}\in\mathbb{R}$, $j=1,2,\ldots, M_i$ such that
\begin{equation}
\mathcal{X}_{i,goal}=\{p\in\mathbb{R}^2|a_{i,j}^Tp+b_{i,j}\le 0, j=1,2,\ldots, M_i\}.
\end{equation}

In other words,
\begin{equation}
\mathcal{X}_{i,goal}=\{x\in\mathcal{X}|a_{i,j}^T[I_2\quad O_2]x+b_{i,j}\le 0, j=1,2,\ldots, M_i\}.
\end{equation}
where $I_2, O_2\in\mathbb{R}^{2\times 2}$ denote the 2-dimensional identity and zero matrices, respectively.


We assume that all agents share a synchronized clock. The terminal time of multi-agent motion is upper-bounded by $t_f=T_f\Delta t$ with $T_f\in\mathbb{N}^+$, and the planning horizon is then given by $[0,T_f]$. Accomplishment of individually-assigned specifications is of practical importance, for instance search and rescue missions or coverage tasks are often specified to mobile robots individually. In this paper, local motion planning tasks for agent $i$ are summarized as the following STL formula: for $i\in\mathcal{P}$, require:

\begin{equation}
\varphi_i=\varphi_{i,p}\land\varphi_{i,s},
\label{stl}
\end{equation}
where
\begin{enumerate}
\item the motion performance property
\begin{equation}
\varphi_{i,p}=\Diamond_{[0,T_f]} \bigwedge_{j=1}^{M_i}\left(a_{i,j}^T[I_2\quad O_2]x_i+b_{i,j}\le 0\right)
\end{equation}
requires that agent $i$ enter the goal region within $T_f$ time steps;
\item the safety property
\begin{equation}
\begin{aligned}
\varphi_{i,s} &= \Box_{[0,T_f]} \bigwedge_{j\in\mathcal{N}_i, j\ne i}[(|p_{i,1}-p_{j,1}|\ge d_1)\\
&\land(|p_{i,2}-p_{j,2}|\ge d_2)]
\end{aligned}
\end{equation}
ensures that agent $i$ shall never encounter obstacle regions. Here $d_1$ and $d_2$ are pre-defined safety distances between two agents in the two dimensions. $\mathcal{N}_i \subseteq \mathcal{P}$ denotes the set for agent $i$'s neighbor which will be described in Section V.
\end{enumerate}
\subsection{SpaTeL Specifications of Communication and Safety}
The shared environment $\mathcal{X}$ is divided into $2^D \times 2^D$ same sized grids such that the environment can be represented by a quad tree, where $D$ is the depth of the tree. Without loss of generality, we also assume that the obstacle region $\mathcal{X}_{obs}$, like the mountains in Fig. \ref{comm}, is a rectangular subset of $\mathcal{X}$ which is also represented by grids. Fig. \ref{matrix} and Fig. \ref{tree} illustrate such setup with corresponding QTS, where $D=3$. The number in each grid is the valuation function for each leaf node. We use SpaTeL specifications to specify obstacles avoidance, optimal communication with base stations and preventing traffic congestion. To this regard, we assign each leaf node with a number listed in Fig.\ref{matrix} representing the maximum amount of agents where each grid can have at the same time. For the obstacles represented with red grids, the number of agents allowed in it is zero. We place four communication base stations shown in Fig.~\ref{comm} located at the center of $v_1$ to $v_4$. Considering the path loss and the shadowing effect in wireless communication, we assume grids that are far away from base stations and are blocked by obstacles will suffer poor communication quality leading to the maximum amount of agents in those grids are relatively small. The grids with number 6 are the initial and terminal locations for all agents which have a higher capacity for accommodating agents.

To specify the spatial temporal specifications mentioned above, we formulate the SpaTeL formula as bellow:
\begin{equation}
\phi=\Box_{[0,T_f]}(\psi_1)\land\Box_{[0,T_f]}(\psi_2\land\psi_3\land\psi_4\land\psi_5)
\label{spatel}
\end{equation}
where $\psi_i$ are TSSL formulas:
\begin{equation}
\begin{aligned}
\psi_1=&\forall SW \bigcirc \forall NE \bigcirc \forall\begin{Bmatrix}
NW,NE,SW
\end{Bmatrix} \bigcirc (\mu=0) \land\\
&\forall SE \bigcirc\forall NW \bigcirc\forall NW \bigcirc (\mu=0) \land\\
&\forall NE \bigcirc\forall SW \bigcirc\forall\begin{Bmatrix}
NW,NE
\end{Bmatrix} \bigcirc (\mu=0)
\end{aligned}
\end{equation}
$\psi_1$ specifies the safety property of avoiding obstacles.
\begin{equation}
\begin{aligned}
\psi_2=&\forall NW \bigcirc\forall NW \bigcirc\forall NW \bigcirc (\mu\leq 6) \land\\
&\forall NW \bigcirc \forall NW \bigcirc \forall\begin{Bmatrix}
NE,SW,SE
\end{Bmatrix} \bigcirc (\mu\leq 4) \land\\
&\forall NW \bigcirc \forall NE \bigcirc \forall\begin{Bmatrix}
NW,SE
\end{Bmatrix} \bigcirc (\mu\leq 3) \land\\
&\forall NW \bigcirc\forall NE \bigcirc\forall SW \bigcirc (\mu\leq 4) \land\\
&\forall NW \bigcirc\forall NE \bigcirc\forall NE \bigcirc (\mu\leq 2) \land\\
&\forall NW \bigcirc\forall SW \bigcirc\forall \begin{Bmatrix}
NW,SE
\end{Bmatrix} \bigcirc (\mu\leq 3) \land\\
&\forall NW \bigcirc\forall SW \bigcirc\forall NE \bigcirc (\mu\leq 4) \land\\
&\forall NW \bigcirc\forall SW \bigcirc\forall SW \bigcirc (\mu\leq 2) \land\\
&\forall NW \bigcirc\forall SE \bigcirc\forall NW \bigcirc (\mu\leq 4) \land\\
&\forall NW \bigcirc\forall SE \bigcirc\forall \begin{Bmatrix}
NE,SE
\end{Bmatrix} \bigcirc (\mu\leq 2) \land\\
&\forall NW \bigcirc\forall SE \bigcirc\forall SW \bigcirc (\mu\leq 3)
\end{aligned}
\end{equation}
$\psi_2$ specifies the desired pattern in terms of communication quality and avoiding traffic congestion for the upper left quadrant. $\psi_3$ to $\psi_5$ specify the rest of environment following the same procedure.

Comparing SpaTeL formulas (\ref{spatel}) and STL formulas (\ref{stl}), it is worth pointing out that STL formulas define the requirement for each agent while SpaTeL formulas define global specifications for all agents. By using both STL and SpaTeL formulas in our design, we are able to consider both local and global properties in a decentralized method.
\subsection{MPC based Co-optimization Problem}
We wish to achieve a co-optimization for motion planning and communication QoS. To this regard, the cost function $J_{i,1}$ in the following linear quadratic form represents the energy consumption for agent $i$.

\begin{equation}
J_{i,1}=\sum_{k=k'}^{k'+H-1}(q^T|x_{i,t_k}|+r^T|u_{i,t_k}|)+h(k')\sum_{k=k'}^{k'+H-1}d_{i,k}
\label{J1}
\end{equation}
where $H$ is the planning horizon, $q$ and $r$ are non-negative weighting vectors and $|.|$ denotes the element-wise absolute value. Second term is time penalty multiplying goal penalty such that each agent can move towards its goal. We define goal penalty as $d_{i,k}=||p_{i,k}^T-p_{i,goal}^T||$ and time penalty as $h(k)=\lambda k^2$, where $p_{i,goal} \in \mathbb{R}^2$ denotes the geometric center of goal region $\mathcal{X}_{i,goal}$ for agent $i$ and $\lambda$ is a user defined parameter.

As for communication QoS, we consider both base station-to-agent and agent-to-agent scenarios. For inter-agent communication, we assume in each planning period, for agent $i$, agents within its neighbor are static where the communication channel among them are affected by path loss. For agent $j \in \mathcal{N}_i$, one can defines a matrix $C'_j$ similar to $C$ which defines its communication QoS pattern in the environment. Using similar idea from our previous work \cite{liuacc}, the communication cost is formulated as follow.

\begin{equation}
J_{i,2}=\sum_{k=k'}^{k'+H-1}\sum_{m=1}^{2^D}\sum_{n=1}^{2^D}(C_{m,n}+\sum_{j\in \mathcal{N}_i}^{\mathcal{N}_i}C'_{j,m,n})(O_{m,n,t_k}-1)
\label{J2}
\end{equation}
where $C$ is defined in Fig.\ref{matrix} representing communication quality with base stations. $O_t$ is a binary matrix for capturing the occupancy of the grids where $O_{m,n,t}$ is zero if and only if the agent is in m-th row and n-th column.

Given the aforementioned preliminaries and cost functions, we formally formulate the distributed communication-aware motion planning from STL-SpaTeL specifications as below:
\begin{problem}
	Given a multi-agent system with $P$ agents whose dynamic behaviors are determined by (\ref{agentdynamic}) with initial states $x_{i,0}$, a planning horizon $H$, a local STL formula $\varphi_i$ in (\ref{stl}) and a global formula $\phi$ in (\ref{spatel}), we find local control inputs $u_i(t_k)$ for all agents such that the following cost function is optimized:
	\begin{equation}
	\underset{{\bf u}_i^{H}, i\in\mathcal{P}}{\min}\quad J_i({\bf x}_i(x_{i,0},{\bf u}_i^{H}))=\alpha J_{i,1}+(1-\alpha)J_{i,2}
	\label{J}
	\end{equation}
	\begin{alignat*}{3}
	\mbox{s.t.}\quad &\forall i\in\mathcal{P}, \\
	&x_i(t_{k+1})=A_dx_i(t_k)+B_du_i(t_k),\\
	&{\bf x}_i(x_{i,k},{\bf u}_i^{H}) \models \varphi_i,\\
	&({\bf q},t_k)\models \phi,\\
	&u_i\in\mathcal{U}=[-u_{max}, u_{max}]\times[-u_{max}, u_{max}], \\
	&||v_i||<v_{max},\\
	& \omega_i=\frac{||u_i||}{m_i||v_i||}\leq\frac{u_{max}}{m_iv_{max}}
	\end{alignat*}
where $\alpha \in [0,1]$ is a user defined parameter.	
\end{problem}

\section{MILP Encoding of Communication-aware Motion Planning}

\subsection{MILP Encoding of Agent Dynamics}
In this section, we replace $t_k$ with $t$ and denote $x_{it}$ and $u_{it}$ as the state and control inputs of agent $i$ at time step $t$. To encode the motion planning cost (\ref{J1}) as linear programming, we employ Manhattan distance for $d_{i,k}$ and introduce slack vectors $\alpha_{it}$, $\beta_{it}$, $\gamma_{it}$ and additional constraints \cite{athans2013optimal} such that $J_{i,1}$ can be transformed to linear cost function.
\begin{equation}
J_{i,1}=\sum_{t=k'}^{k'+H-1}({q}^T{\alpha_{it}+{r}^T\beta_{it}})+h(k')\sum_{t=k'}^{k'+H-1}\sum_{k=1}^{2}\gamma_{itk}
\end{equation}

\begin{equation}
\begin{aligned}
&\text{s.t.} & \forall t\in [k',k'+&H-1], \forall j\in [1,4], \forall k\in [1, 2] \\
& \text{}
& x_{itj}&\leq \alpha_{itj}, -x_{itj}\leq \alpha_{itj} \\
& \text{and}
& u_{itk}&\leq \beta_{itk}, -u_{itk}\leq \beta_{itk}\\
& \text{and}
& x_{itk}-p_{i,goal,k}&\leq \gamma_{itk}, -x_{itk}+p_{i,goal,k}\leq \gamma_{itk} \\
& \text{and}
& x_i(t+1)&=A_dx_i(t)+B_du_i(t)
\end{aligned}
\end{equation}

To transfer the nonlinear velocity constraints, we use the method in \cite{richards2002aircraft} by introducing an arbitrary number $L$ of linear constraints leading to the 2-D velocities approximated by a regular L-sided polygon.
\begin{equation}
\begin{split}
\forall l\in [1,L], i\in[1,P],t\in[k',k'+H-1]\\
v_{it1}sin(\frac{2\pi l}{L})+v_{it2}cos(\frac{2\pi l}{L})\leq v_{max}
\end{split}
\end{equation}
\subsection{Boolean Encoding of STL Constraints}
For the MILP encoding of STL specifications in (\ref{stl}), we denote two Boolean variables $z_{t}^{\varphi_{i,p}}$ and $z_{t}^{\varphi_{i,s}}$ whose value depends on the satisfaction of $\varphi_{i,p}$ and $\varphi_{i,s}$ respectively  \cite{raman2014model,liuacc}. Then the satisfaction of $\varphi_i$ at time step $t$ can be represented by Boolean variable $z_{t}^{\varphi_i}$ which is determined by $z_{t}^{\varphi_{i,p}}$ and $z_{t}^{\varphi_{i,s}}$ as follow.

\begin{equation}
z_{t}^{\varphi_i}=z_{t}^{\varphi_{i,p}} \land z_{t}^{\varphi_{i,s}}
\end{equation}
with
\begin{equation}
\begin{split}
\forall i\in\mathcal{P}: \\
&z_{t}^{\varphi_i}\le z_{t}^{\varphi_{i,p}}, z_{t}^{\varphi_i}\le z_{t}^{\varphi_{i,s}}\\
&z_{t}^{\varphi_i}\ge z_{t}^{\varphi_{i,p}}+z_{t}^{\varphi_{i,s}}-1
\end{split}
\end{equation}
where $z_{t}^{\varphi_{i,p}}$ and $z_{t}^{\varphi_{i,s}}$ are one if and only if their corresponding specifications are satisfied.

\subsection{Boolean Encoding of SpaTeL Constraints}
Similar to encoding STL as MILP, we use the method in \cite{haghighi2016robotic} to encode SpaTeL formulas (\ref{spatel}) as MILP recursively. We denote SpaTeL formulas $\phi_1=\Box_{[0,T_f]}(\psi_1)$ and $\phi_2=\Box_{[0,T_f]}(\psi_2\land\psi_3\land\psi_4\land\psi_5)$. We define three Boolean variables  $z_{v,t}^{\phi}$, $z_{v,t}^{\phi_1}$ and $z_{v,t}^{\phi_2}$ whose truth values correspond to the satisfaction of $\phi$, $\phi_1$ and $\phi_2$ respectively and are determined by the following constraints.
\begin{equation}
\begin{split}
&z_{v,t}^{\phi}=z_{v,t}^{\phi_{1}} \land z_{v,t}^{\phi_{2}}\\
&z_{v,t}^{\phi_{1}}=\bigwedge_{k=t}^{T_f} z_{v,k}^{\psi_1},
z_{v,t}^{\phi_{2}}=\bigwedge_{k=t}^{T_f} z_{v,k}^{\psi_{2-5}} \\
&z_{v,k}^{\psi_{2-5}}=z_{v,k}^{\psi_{2}} \land z_{v,k}^{\psi_{3}} \land z_{v,k}^{\psi_{4}} \land z_{v,k}^{\psi_{5}}
\end{split}
\end{equation}
with
\begin{equation}
\begin{split}
&\forall i\in\mathcal{P}, \forall k\in [t,T_f]: \\
&z_{v,t}^{\phi}\le z_{v,t}^{\phi_{1}}, z_{v,t}^{\phi}\le z_{v,t}^{\phi_{2}}, z_{v,t}^{\phi}\ge z_{v,t}^{\phi_{1}}+z_{v,t}^{\phi_{2}}-1\\
&z_{v,t}^{\phi_1}\le z_{v,k}^{\psi_{1}}, z_{v,t}^{\phi_2}\le z_{v,k}^{\psi_{2-5}}\\
&z_{v,t}^{\phi_1}\ge \sum_{k=t}^{T_f}z_{v,k}^{\psi_{1}}-T_f+t, z_{v,t}^{\phi_2}\ge \sum_{k=t}^{T_f}z_{v,k}^{\psi_{2-5}}-T_f+t \\
&z_{v,k}^{\psi_{2-5}}\le z_{v,k}^{\psi_{2}},z_{v,k}^{\psi_{2-5}}\le z_{v,k}^{\psi_{3}},z_{v,k}^{\psi_{2-5}}\le z_{v,k}^{\psi_{4}},z_{v,k}^{\psi_{2-5}}\le z_{v,k}^{\psi_{5}}\\
&z_{v,k}^{\psi_{2-5}}\ge \sum_{j=2}^{5}z_{v,k}^{\psi_{j}}-3 \\
\end{split}
\end{equation}
where Boolean variables $z_{v,k}^{\psi_{i}}$, $i\in[1,\dots,5]$ represent the satisfaction of TSSL formulas $\psi_1$ to $\psi_5$. We assume the sampling period $\Delta t$ for the discretized system is 1.
\section{Distributed Model Predictive Control Synthesis}
We wish to construct a distributed and online framework for the communication-aware motion planning. Towards this end, we employ MPC as the basic framework such that the sub-problem for each agent can be solved online and the system is robust enough to deal with model uncertainty and external disturbance. Another strategy we employed for achieving distributed and online manner is only considering its neighbor for each agent during planning so that the size of each sub-problem will reduce significantly for the size of the formulated MILP problem in previous section depends on the number of agents directly. We assume that global information like time, synchronization, states of each agent are available for each agent through communication base stations. The algorithm is summarized in Algorithm \ref{algorithm}.

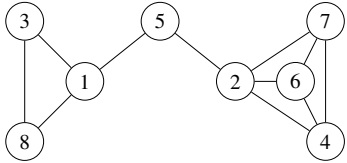
\begin{figure}
	\begin{center}
		\scalebox{0.8}{
			\begin{tikzpicture}
			\node[shape=circle,draw=black] (A) at (0,0) {8};
			\node[shape=circle,draw=black] (B) at (0,2) {3};
			\node[shape=circle,draw=black] (C) at (1,1) {1};
			\node[shape=circle,draw=black] (D) at (2.25,2) {5};
			\node[shape=circle,draw=black] (F) at (3.5,1) {2};
			\node[shape=circle,draw=black] (E) at (4.5,1) {6};
			\node[shape=circle,draw=black] (G) at (5,2) {7};
			\node[shape=circle,draw=black] (H) at (5,0) {4};
			
			\path [-] (A) edge node[left] {} (B);
			\path [-](B) edge node[left] {} (C);
			\path [-](A) edge node[left] {} (C);
			\path [-](D) edge node[left] {} (C);
			\path [-](D) edge node[above] {} (F);
			\path [-](E) edge node[above] {} (F);
			\path [-](E) edge node[above] {} (G);
			\path [-](G) edge node[above] {} (H);
			\path [-](E) edge node[above] {} (H);
			\path [-](F) edge node[above] {} (G);
			\path [-](F) edge node[above] {} (H);
			\end{tikzpicture}
		}
	\end{center}
	\caption{Neighbor}
	\label{neighbor}
\end{figure}

Since it is reasonable to omit the agents that are far away, we define the neighbor of each agent by choosing the agents whose distance to the agent are shorter than a certain threshold. We assign each agent a unique priority in each planning period randomly. Agents can only plan after all the agents in its neighbor with higher priority have planned. Fig. \ref{neighbor} illustrates our idea. Agents connected with edges are considered in the same neighbor. For the agent with priority 2, it plans first in each period since it has the highest priority in its neighbor and the agent with priority 4 will plan right after it.

In each planning period, agents formulate their own optimization problem in (\ref{J}) encoded as MILP and run commercial MILP solver to find the control inputs within the planning horizon. After all agents planned, they implement the first step of the control inputs and move into next period. The algorithm stops when all agents reach their goals or it reaches the time limit.
\begin{algorithm}
	\SetAlgoLined
	\KwIn{Intial states and goal positions of each agent}
	
	\KwOut{Return states and control inputs of each agent}
	
	Initialization

	\While{AgentSet $\neq \emptyset$ or $t \leq T_f$}{
		Randomly set a unique priority for each agent in AgentSet
		
		\For{Agent $i$, at time $t$}{
			Update the list of all agents' states  ${\bf x}_j^H$ and control inputs ${\bf u}_j^H$, $j\in\mathcal{P}$
			
			\If{Agent $j$, $j\in\mathcal{P}$ is close enough to agent $i$}{
				Add agent $j$ into agent $i$'s neighbour $\mathcal{N}_i$
			}
			
			Wait until all agents in $\mathcal{N}_i$ with higher priority than $i$ planned, then
			
			\For{$t\in[t,t+H]$}{
				Minimize the cost function $J_i$ in (\ref{J}) subjecting to corresponding constraints
				}
			
			Broadcast the results of ${\bf x}_i^H$ and  ${\bf u}_i^H$ to the whole group
			}
			
			Implement the first step of control inputs ${\bf u}_i^H$
			
			Global time $t=t+1$
			
		\If{Any agent $i\in\mathcal{P}$ reaches its target}{
			Remove it from AgentSet
		}
	}
\caption{Distributed Communication-aware Motion Planning with MPC}
\label{algorithm}
\end{algorithm}

\section{Simulation Results}
To justify our distributed co-optimization strategy, we implemented our MPC based framework with MATLAB. The encoded MILP problem was modeled by AMPL, an algebraic modeling language for large scale mathematical programming \cite{fourer1993ampl}, and solved by Gurobi, a commercial solver for MILP \cite{optimization2012gurobi}.

\begin{figure}[h]
	\centering
	\includegraphics[width=1.1\linewidth]{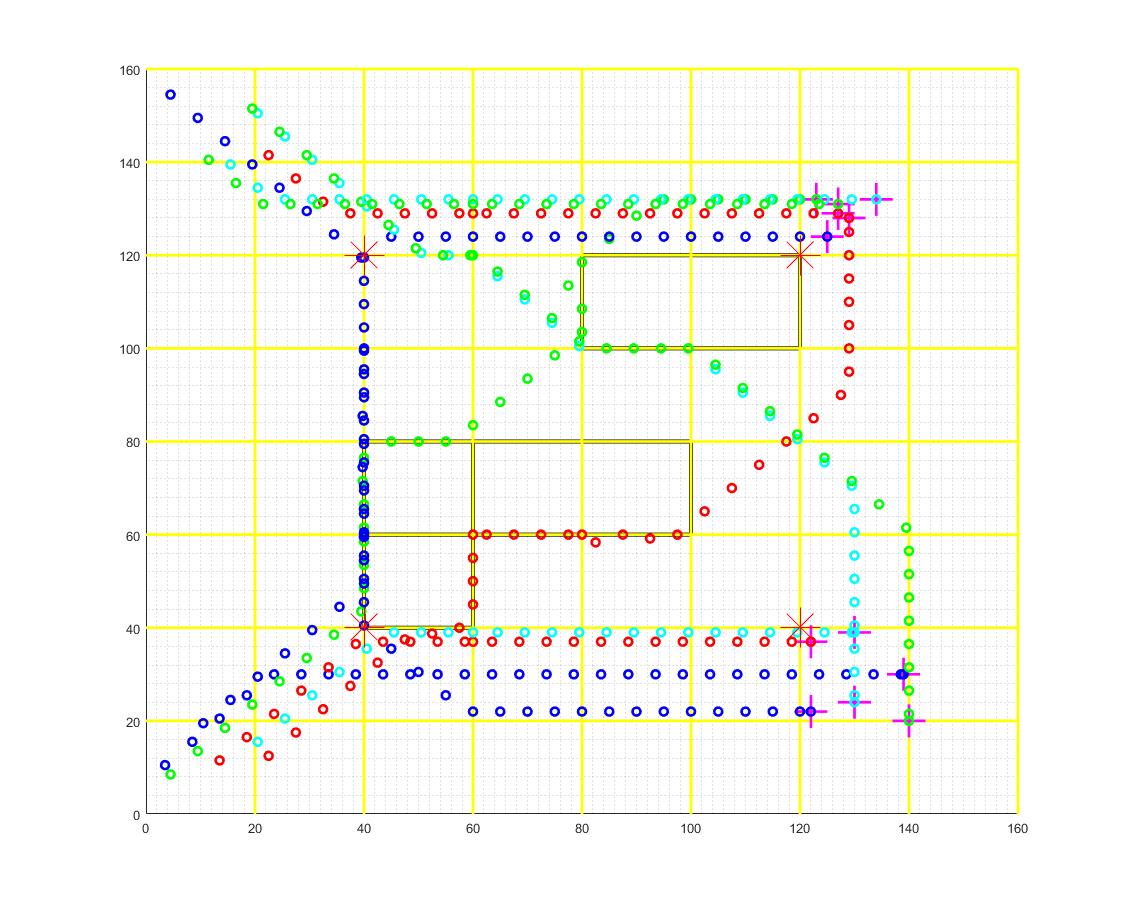}
	\caption{Distributed Communication-aware Motion Planning}
	\label{Data2}
\end{figure}

\begin{figure}[h]
	\centering
	\includegraphics[width=1.1\linewidth]{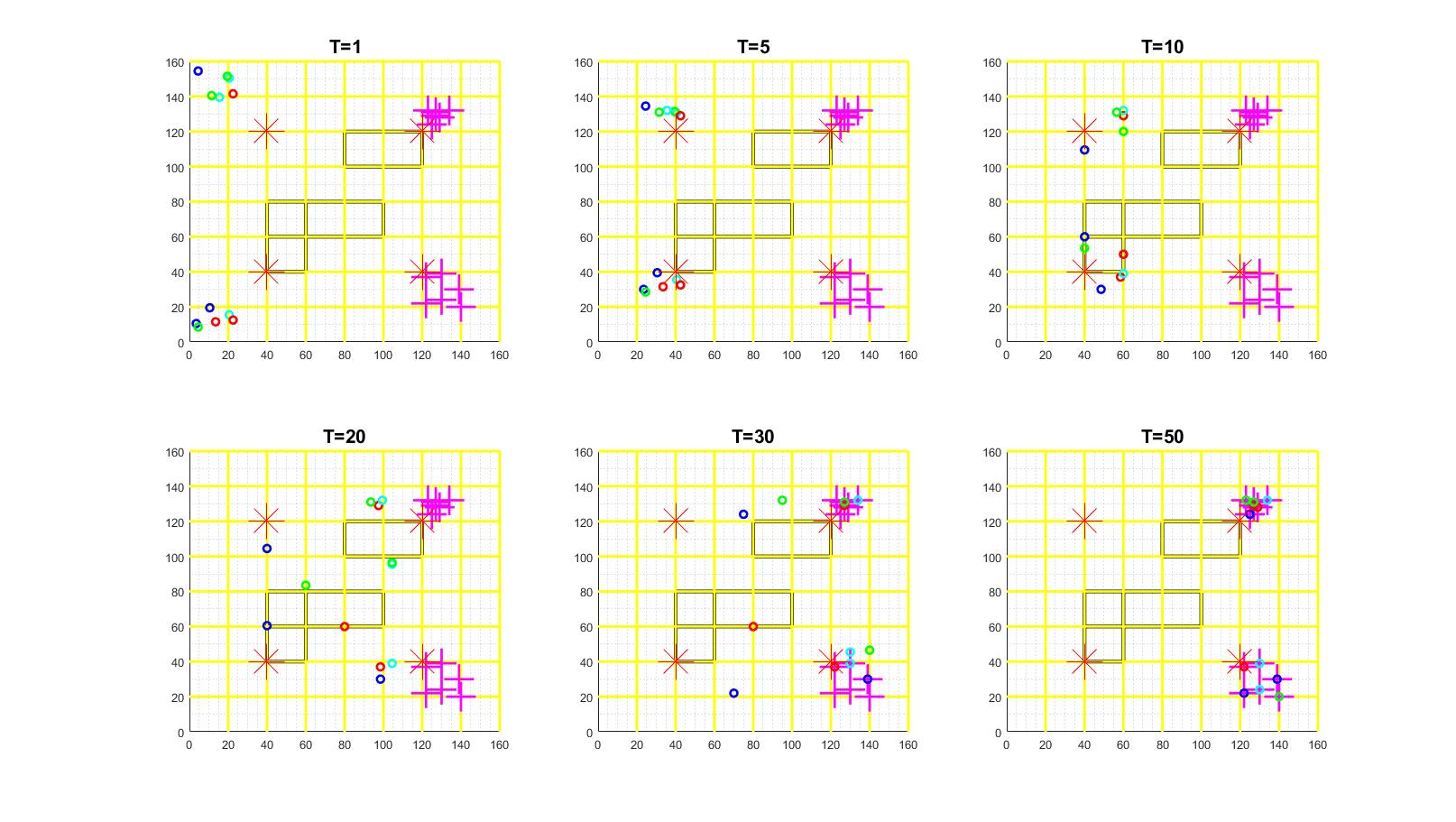}
	\caption{Agents distribution}
	\label{Data3}
\end{figure}

Fig. \ref{comm} illustrates our basic setup. Fig. \ref{Data2} shows the simulation result of the MPC based distributed co-optimization. Fig. \ref{Data3} demonstrates agents distribution at different time steps. Twelve agents with initial states and target positions marked as red cross are given. The agent dynamics ruled by (\ref{agentdynamic}) is given by setting matrices $A_d$ and $B_d$ as:
\begin{equation}
A_d=
\begin{bmatrix}
1 &0 &1 &0 \\
0 &1 &0 &1 \\
0 &0 &1 &0 \\
0 &0 &0 &1
\end{bmatrix},
B_d=
\begin{bmatrix}
0.5 &0\\
0 &0.5\\
1 &0\\
0 &1
\end{bmatrix}.
\end{equation}
We choose $H=5$, $L=8$, $\Delta t=1$, $T_f=50$, $\lambda =0.005$, $d_1=d_2=1$, $\alpha=0.5$. The working space is set as a $160m \times 160m$ square which is represented by QTS with $D=3$ shown in yellow grids. Four communication base stations are marked as red stars in the graph. Obstacles are marked as black rectangular. Communication quality is considered in this case. Due to the shadowing effect caused by obstacles, the communication quality is the worst in the space between two obstacles. Fig. \ref{matrix} shows the communication quality pattern in terms of base stations and obstacles. The outputs of the simulation are agents' states and control inputs at all steps.

\begin{figure}[h]
	\centering
	\includegraphics[width=1.1\linewidth]{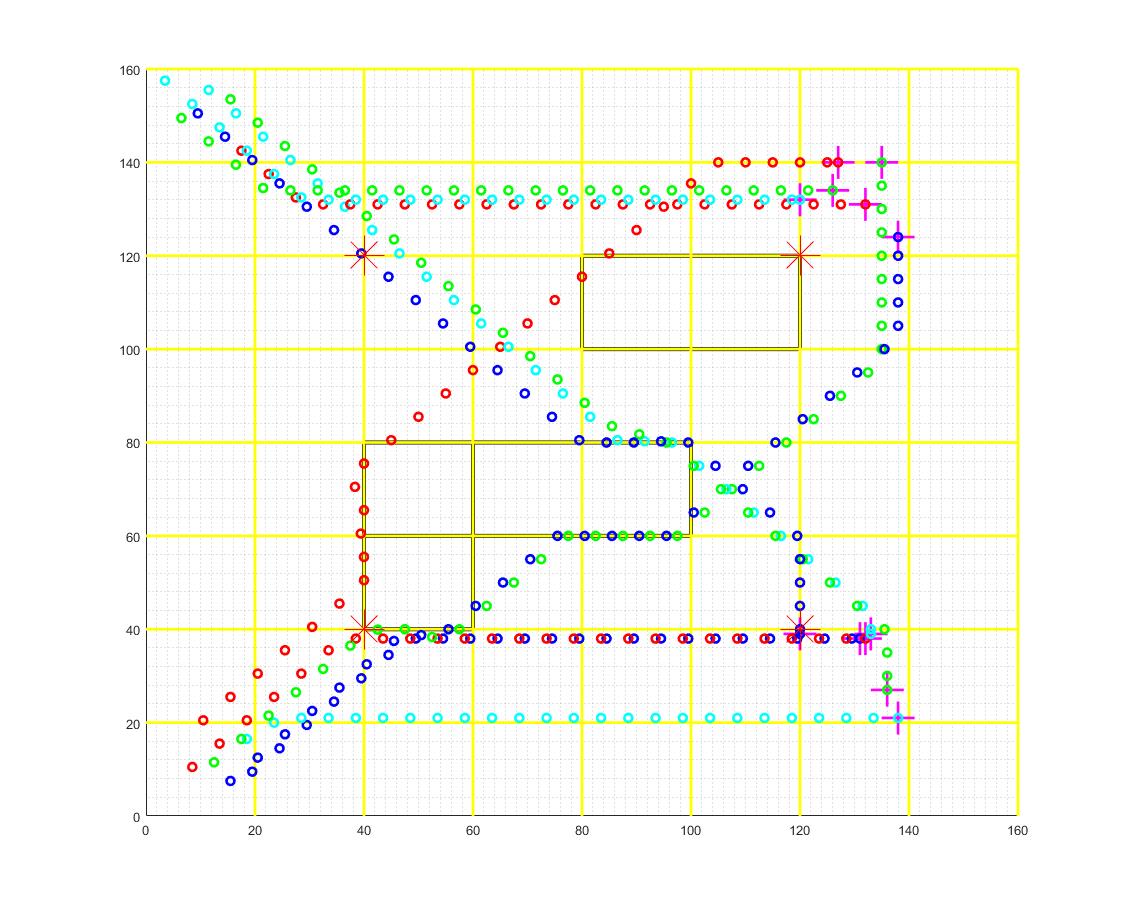}
	\caption{Distributed Motion Planning}
	\label{Data1}
\end{figure}

As we can see from the result, using the proposed distributed co-optimization strategy, all agents are able to satisfy the specifications. Several agents avoid the poor communication quality area to reach their goals, compared to Fig. \ref{Data1} without considering communication.

The simulation was run on a PC with Intel core i7-4710MQ 2.50 GHz processor and 8GB RAM. The algorithm was run distributively among agents. Each agent can solve its own MILP problem around 0.1 second.

\section{Conclusion}
As an extension of our previous work \cite{liuacc}, we develop a distributed MPC based algorithm for communication-aware motion planning. By using STL-SpaTeL formulas to specify motion planning and communication requirements and encoding them into MILP under distributed MPC, the proposed algorithm is able to find online control inputs for each agent distributively such that desired specifications and patterns can be satisfied and hence demonstrates the ability of dealing with large scale systems. The algorithm is validated by a co-optimization simulation for multi-agent system.

\bibliographystyle{IEEEtran}
\bibliography{comm}

\begin{thebibliography}{10}
\providecommand{\url}[1]{#1}
\csname url@samestyle\endcsname
\providecommand{\newblock}{\relax}
\providecommand{\bibinfo}[2]{#2}
\providecommand{\BIBentrySTDinterwordspacing}{\spaceskip=0pt\relax}
\providecommand{\BIBentryALTinterwordstretchfactor}{4}
\providecommand{\BIBentryALTinterwordspacing}{\spaceskip=\fontdimen2\font plus
\BIBentryALTinterwordstretchfactor\fontdimen3\font minus
  \fontdimen4\font\relax}
\providecommand{\BIBforeignlanguage}[2]{{%
\expandafter\ifx\csname l@#1\endcsname\relax
\typeout{** WARNING: IEEEtran.bst: No hyphenation pattern has been}%
\typeout{** loaded for the language `#1'. Using the pattern for}%
\typeout{** the default language instead.}%
\else
\language=\csname l@#1\endcsname
\fi
#2}}
\providecommand{\BIBdecl}{\relax}
\BIBdecl

\bibitem{cao2013overview}
Y.~Cao, W.~Yu, W.~Ren, and G.~Chen, ``An overview of recent progress in the
  study of distributed multi-agent coordination,'' \emph{IEEE Transactions on
  Industrial informatics}, vol.~9, no.~1, pp. 427--438, 2013.

\bibitem{baier2008principles}
C.~Baier, J.-P. Katoen, and K.~G. Larsen, \emph{Principles of model
  checking}.\hskip 1em plus 0.5em minus 0.4em\relax Boston: MIT Press, 2008.

\bibitem{kloetzer2008fully}
M.~Kloetzer and C.~Belta, ``A fully automated framework for control of linear
  systems from temporal logic specifications,'' \emph{IEEE Transactions on
  Automatic Control}, vol.~53, no.~1, pp. 287--297, 2008.

\bibitem{belta2007symbolic}
C.~Belta, A.~Bicchi, M.~Egerstedt, E.~Frazzoli, E.~Klavins, and G.~J. Pappas,
  ``Symbolic planning and control of robot motion [grand challenges of
  robotics],'' \emph{IEEE Robotics \& Automation Magazine}, vol.~14, no.~1, pp.
  61--70, 2007.

\bibitem{kloetzer2010automatic}
M.~Kloetzer and C.~Belta, ``Automatic deployment of distributed teams of robots
  from temporal logic motion specifications,'' \emph{IEEE Transactions on
  Robotics}, vol.~26, no.~1, pp. 48--61, 2010.

\bibitem{fainekos2009temporal}
G.~E. Fainekos, A.~Girard, H.~Kress-Gazit, and G.~J. Pappas, ``Temporal logic
  motion planning for dynamic robots,'' \emph{Automatica}, vol.~45, no.~2, pp.
  343--352, 2009.

\bibitem{karaman2009sampling}
S.~Karaman and E.~Frazzoli, ``Sampling-based motion planning with deterministic
  $\mu$-calculus specifications,'' in \emph{Decision and Control, 2009 held
  jointly with the 2009 28th Chinese Control Conference. CDC/CCC 2009.
  Proceedings of the 48th IEEE Conference on}.\hskip 1em plus 0.5em minus
  0.4em\relax IEEE, 2009, pp. 2222--2229.

\bibitem{richards2002aircraft}
A.~Richards and J.~P. How, ``Aircraft trajectory planning with collision
  avoidance using mixed integer linear programming,'' in \emph{Proceedings of
  the 2002 American Control Conference}, vol.~3.\hskip 1em plus 0.5em minus
  0.4em\relax IEEE, 2002, pp. 1936--1941.

\bibitem{karaman2011linear}
S.~Karaman and E.~Frazzoli, ``Linear temporal logic vehicle routing with
  applications to multi-uav mission planning,'' \emph{International Journal of
  Robust and Nonlinear Control}, vol.~21, no.~12, pp. 1372--1395, 2011.

\bibitem{wolff2014optimization}
E.~M. Wolff, U.~Topcu, and R.~M. Murray, ``Optimization-based trajectory
  generation with linear temporal logic specifications,'' in \emph{2014 IEEE
  International Conference on Robotics and Automation (ICRA)}.\hskip 1em plus
  0.5em minus 0.4em\relax IEEE, 2014, pp. 5319--5325.

\bibitem{wongpiromsarn2012receding}
T.~Wongpiromsarn, U.~Topcu, and R.~M. Murray, ``Receding horizon temporal logic
  planning,'' \emph{IEEE Transactions on Automatic Control}, vol.~57, no.~11,
  pp. 2817--2830, 2012.

\bibitem{tumova2016multi}
J.~Tumova and D.~V. Dimarogonas, ``Multi-agent planning under local ltl
  specifications and event-based synchronization,'' \emph{Automatica}, vol.~70,
  pp. 239--248, 2016.

\bibitem{kuwata2011cooperative}
Y.~Kuwata and J.~P. How, ``Cooperative distributed robust trajectory
  optimization using receding horizon milp,'' \emph{IEEE Transactions on
  Control Systems Technology}, vol.~19, no.~2, pp. 423--431, 2011.

\bibitem{grancharova2015uavs}
A.~Grancharova, E.~I. Gr{\o}tli, D.-T. Ho, and T.~A. Johansen, ``Uavs
  trajectory planning by distributed mpc under radio communication path loss
  constraints,'' \emph{Journal of Intelligent \& Robotic Systems}, vol.~79,
  no.~1, pp. 115--134, 2015.

\bibitem{grotli2012path}
E.~I. Gr{\o}tli and T.~A. Johansen, ``Path planning for uavs under
  communication constraints using splat! and milp,'' \emph{Journal of
  Intelligent \& Robotic Systems}, vol.~65, no. 1-4, pp. 265--282, 2012.

\bibitem{yan2013co}
Y.~Yan and Y.~Mostofi, ``Co-optimization of communication and motion planning
  of a robotic operation under resource constraints and in fading
  environments,'' \emph{IEEE Transactions on Wireless Communications}, vol.~12,
  no.~4, pp. 1562--1572, 2013.

\bibitem{maler2004monitoring}
O.~Maler and D.~Nickovic, ``Monitoring temporal properties of continuous
  signals,'' in \emph{Formal Techniques, Modelling and Analysis of Timed and
  Fault-Tolerant Systems}.\hskip 1em plus 0.5em minus 0.4em\relax Springer,
  2004, pp. 152--166.

\bibitem{raman2014model}
V.~Raman, A.~Donz{\'e}, M.~Maasoumy, R.~M. Murray, A.~Sangiovanni-Vincentelli,
  and S.~A. Seshia, ``Model predictive control with signal temporal logic
  specifications,'' in \emph{Proceedings of the 53rd IEEE Conference on
  Decision and Control (CDC)}.\hskip 1em plus 0.5em minus 0.4em\relax IEEE,
  2014, pp. 81--87.

\bibitem{haghighi2016robotic}
I.~Haghighi, S.~Sadraddini, and C.~Belta, ``Robotic swarm control from
  spatio-temporal specifications,'' in \emph{Decision and Control (CDC), 2016
  IEEE 55th Conference on}.\hskip 1em plus 0.5em minus 0.4em\relax IEEE, 2016,
  pp. 5708--5713.

\bibitem{liuacc}
Z.~Liu, J.~Dai, B.~Wu, and H.~Lin, ``Communication-aware motion planning for
  multi-agent systems from signal temporal logic specifications,'' in
  \emph{American Control Conference (ACC), Proceedings of the 2017}.

\bibitem{raman2015reactive}
V.~Raman, A.~Donz{\'e}, D.~Sadigh, R.~M. Murray, and S.~A. Seshia, ``Reactive
  synthesis from signal temporal logic specifications,'' in \emph{Proceedings
  of the 18th International Conference on Hybrid Systems: Computation and
  Control (HSCC)}.\hskip 1em plus 0.5em minus 0.4em\relax ACM, 2015, pp.
  239--248.

\bibitem{molisch2012wireless}
A.~F. Molisch, \emph{Wireless communications}.\hskip 1em plus 0.5em minus
  0.4em\relax John Wiley \& Sons, 2012, vol.~34.

\bibitem{haghighi2015spatel}
I.~Haghighi, A.~Jones, Z.~Kong, E.~Bartocci, R.~Gros, and C.~Belta, ``Spatel: a
  novel spatial-temporal logic and its applications to networked systems,'' in
  \emph{Proceedings of the 18th International Conference on Hybrid Systems:
  Computation and Control}.\hskip 1em plus 0.5em minus 0.4em\relax ACM, 2015,
  pp. 189--198.

\bibitem{bartocci2016formal}
E.~Bartocci, E.~A. Gol, I.~Haghighi, and C.~Belta, ``A formal methods approach
  to pattern recognition and synthesis in reaction diffusion networks,''
  \emph{IEEE Transactions on Control of Network Systems}, 2016.

\bibitem{mayne2000constrained}
D.~Q. Mayne, J.~B. Rawlings, C.~V. Rao, and P.~O. Scokaert, ``Constrained model
  predictive control: Stability and optimality,'' \emph{Automatica}, vol.~36,
  no.~6, pp. 789--814, 2000.

\bibitem{kuwata2007distributed}
Y.~Kuwata, A.~Richards, T.~Schouwenaars, and J.~P. How, ``Distributed robust
  receding horizon control for multivehicle guidance,'' \emph{IEEE Transactions
  on Control Systems Technology}, vol.~15, no.~4, pp. 627--641, 2007.

\bibitem{athans2013optimal}
M.~Athans and P.~L. Falb, \emph{Optimal control: an introduction to the theory
  and its applications}.\hskip 1em plus 0.5em minus 0.4em\relax Courier
  Corporation, 2013.

\bibitem{fourer1993ampl}
R.~Fourer, D.~Gay, and B.~Kernighan, \emph{Ampl}.\hskip 1em plus 0.5em minus
  0.4em\relax Boyd \& Fraser Danvers, MA, 1993, vol. 117.

\bibitem{optimization2012gurobi}
Z.~Gu, E.~Rothberg, and R.~Bixby, ``Gurobi optimizer reference manual,''
  \emph{URL: http://www. gurobi. com}, vol.~2, pp. 1--3, 2012.

\end{thebibliography}
\end{document}